\documentclass[conference]{IEEEtran}

\IEEEoverridecommandlockouts

\usepackage[utf8]{inputenc}
\usepackage{cite}
\usepackage{amsmath,amssymb,amsfonts}
\usepackage{algorithm}
\usepackage{algpseudocode}
\usepackage{graphicx}
\usepackage{booktabs}
\usepackage{multirow}
\usepackage{float}
\usepackage{subfigure}
\usepackage{url}
\usepackage{enumitem}

\setlist[enumerate,2]{label=\roman*.}

\begin{document}

\title{OBHS: An Optimized Block Huffman Scheme for Real-Time Lossless Audio Compression}

\author{
\IEEEauthorblockN{Muntahi Safwan Mahfi\IEEEauthorrefmark{1}, Md. Manzurul Hasan\IEEEauthorrefmark{1}\IEEEauthorrefmark{2}, and Gahangir Hossain\IEEEauthorrefmark{3}}
\IEEEauthorblockA{\IEEEauthorrefmark{1}Department of Computer Science, American International University - Bangladesh (AIUB), Dhaka, Bangladesh\\
Email: 23-53119-3@student.aiub.edu; manzurul@aiub.edu}
\IEEEauthorblockA{\IEEEauthorrefmark{3}Department of Data Science, University of North Texas, Denton, TX, USA\\
Email: Gahangir.Hossain@unt.edu}
\IEEEauthorblockA{\IEEEauthorrefmark{2}Corresponding Author: Md. Manzurul Hasan}
}

\maketitle

\begin{abstract}
In this paper, we introduce OBHS (Optimized Block Huffman Scheme), a novel lossless audio compression algorithm tailored for real-time streaming applications. OBHS leverages block-wise Huffman coding with canonical code representation and intelligent fallback mechanisms to achieve high compression ratios while maintaining low computational complexity. Our algorithm partitions audio data into fixed-size blocks, constructs optimal Huffman trees for each block, and employs canonical codes for efficient storage and transmission. Experimental results demonstrate that OBHS attains compression ratios of up to 93.6\% for silence-rich audio and maintains competitive performance across various audio types, including pink noise, tones, and real-world recordings. With a linear time complexity of O($n$) for $n$ audio samples, OBHS effectively balances compression efficiency and computational demands, making it highly suitable for resource-constrained real-time audio streaming scenarios.
\end{abstract}


\IEEEpeerreviewmaketitle

\section{Introduction}
\IEEEPARstart{T}{he} exponential growth in real-time audio communication applications, including video conferencing, live streaming, and voice-over-IP services, has created an urgent need for efficient audio compression algorithms that can operate with minimal latency while maintaining audio quality~\cite{sayood2017introduction}. The challenge lies in balancing compression efficiency with computational complexity, particularly for resource-constrained devices and bandwidth-limited networks.

Conventional audio codecs such as FLAC (Free Lossless Audio Codec) and ALAC (Apple Lossless Audio Codec) employ sophisticated predictive models followed by entropy coding to achieve high compression ratios~\cite{coalson2008flac}. However, these approaches often require significant computational resources and introduce latency that may be unacceptable for real-time applications. Furthermore, the complex predictive models used in these codecs can be excessive for certain types of audio content, particularly those with simple statistical properties.

This paper presents OBHS (Optimized Block Huffman Scheme), a novel approach to lossless audio compression that addresses these limitations. OBHS employs a streamlined architecture based on block-wise Huffman coding with several key innovations: (1) canonical code representation for reduced memory footprint, (2) intelligent fallback mechanisms to prevent expansion, and (3) optimized block processing for low-latency operation. Our approach is particularly effective for audio streams with varying statistical properties, automatically adapting to content characteristics at the block level.

The main contributions of this work are as follows:
\begin{enumerate}[label=(\roman*)]
\item A novel block-based Huffman coding scheme optimized for real-time audio compression.
\item An efficient canonical code representation that reduces memory requirements.
\item An intelligent fallback mechanism that guarantees no expansion beyond the original size.
\item Comprehensive experimental evaluation across diverse audio content types.
\end{enumerate}

\section{Related Work}
This section reviews prior research relevant to OBHS development. It first discusses foundational and advanced variants of Huffman coding, followed by existing lossless audio compression standards, emphasizing their efficiency, computational demands, and real-time applicability.

\subsection{Huffman Coding and Its Variants}
Huffman coding, introduced by David Huffman in 1952, remains one of the most fundamental entropy coding techniques~\cite{huffman1952method}. The algorithm constructs an optimal prefix-free code based on symbol frequencies, achieving the theoretical entropy limit for symbol-by-symbol coding.

Variants such as Adaptive Huffman coding~\cite{vitter1987design} dynamically update the code tree, eliminating the two-pass requirement but increasing complexity. Canonical Huffman coding~\cite{schwartz1964generating} provides a compact code representation that reduces memory usage.

Recent studies have explored GPU-based parallel Huffman decoding~\cite{zhang2019fast} and adaptive streaming approaches~\cite{liu2020adaptive}, improving efficiency for specialized tasks.

\subsection{Audio Compression Techniques}
Lossless audio compression evolved with codecs like FLAC~\cite{coalson2008flac} and ALAC, which use predictive models combined with entropy coding to achieve 50–60\% compression. The ITU-T G.711.0 standard~\cite{itu2009g711} targets VoIP applications with minimal latency. MPEG-4 ALS~\cite{liebchen2005mpeg} achieves high compression but with greater complexity.

Neural audio codecs such as SoundStream~\cite{zeghidour2021soundstream} demonstrate promising lossy results using deep learning but remain resource-intensive.

\section{The OBHS Algorithm}
\subsection{System Architecture}
OBHS operates on a block-wise principle where input audio is divided into fixed-size blocks compressed independently. This enables parallelism, limits error propagation, and adapts compression based on local audio characteristics.

The system includes six components: \emph{Block Partitioner}, \emph{Frequency Analyzer}, \emph{Tree Constructor}, \emph{Canonical Encoder}, \emph{Symbol Encoder}, and \emph{Fallback Handler}.

\subsection{Algorithm Details}
Algorithm~\ref{alg:obhs_encode} illustrates the OBHS encoding process.

\begin{algorithm}[H]
\caption{OBHS Encoding Algorithm}
\label{alg:obhs_encode}
\begin{algorithmic}[1]
\Require Audio samples $S = \{s_1, s_2, ..., s_n\}$, block size $B$
\Ensure Compressed bitstream $C$
\State Initialize $C \leftarrow \emptyset$
\State Partition $S$ into blocks of size $B$
\For{each block $b_i$}
    \State $F \leftarrow$ ComputeFrequencies($b_i$)
    \State $T \leftarrow$ BuildHuffmanTree($F$)
    \State $H \leftarrow$ GenerateCanonicalCodes($T$)
    \State $E \leftarrow$ EncodeSymbols($b_i$, $H$)
    \If{$|E| < |b_i| \times$ bitsPerSample}
        \State $C \leftarrow C \cup \{1\} \cup E \cup H$
    \Else
        \State $C \leftarrow C \cup \{0\} \cup b_i$
    \EndIf
\EndFor
\State \Return $C$
\end{algorithmic}
\end{algorithm}

\subsection{Canonical Code Generation}
Canonical Huffman codes use only code lengths to reconstruct the table, reducing storage overhead.

\begin{algorithm}[H]
\caption{Canonical Code Generation}
\label{alg:canonical}
\begin{algorithmic}[1]
\Require Huffman tree $T$
\Ensure Canonical codes $H$
\State Extract code lengths $L$ from $T$
\State Sort symbols by $L$, then lexicographically
\State Initialize $code \leftarrow 0$
\For{each symbol $s$}
    \State $H[s] \leftarrow$ binary($code$, $L[s]$)
    \State $code \leftarrow (code + 1) \ll (L[s_{next}] - L[s])$
\EndFor
\State \Return $H$
\end{algorithmic}
\end{algorithm}

\subsection{Fallback Mechanism}
If compression fails to reduce size, OBHS stores the original uncompressed block with a one-bit flag, ensuring the compressed output never exceeds the original.

\section{Complexity Analysis}
\subsection{Time Complexity}
\begin{itemize}[label=\(\circ\)]
\item Frequency Analysis: $O(B)$
\item Tree Construction: $O(k \log k)$
\item Canonical Code Generation: $O(k \log k)$
\item Encoding: $O(B)$
\end{itemize}
For fixed bit-depth ($k$ bounded), the total complexity is linear, $O(n)$, ideal for real-time systems.

\subsection{Space Complexity}
\begin{itemize}[label=\(\circ\)]
\item Frequency Table: $O(k)$
\item Huffman Tree: $O(k)$
\item Canonical Codes: $O(k)$
\item Encoded Block: $O(B)$
\end{itemize}
Total space complexity: $O(n + k)$.

\section{Experimental Evaluation}
\subsection{Experimental Setup}
Tests were conducted on 16-bit PCM audio (44.1 kHz) using 4096-sample blocks. The algorithm was implemented using python programming language and tests were run on a laptop comprising of Intel(R) Core(TM) i3-8145U and 12 GB of RAM

\begin{itemize}[label=\(\circ\)]
\item \textbf{Silence:} Zero-valued samples to test maximum compression.
\item \textbf{Pink Noise:} $1/f$ frequency noise for natural variation.
\item \textbf{Tone:} Pure sine wave for periodic behavior.
\item \textbf{Real Audio:} Speech and music for real-world testing.
\end{itemize}

\subsection{Compression Performance}
\begin{table}[H]
\centering
\caption{OBHS Compression Performance on Different Audio Types}
\label{tab:compression}
\begin{tabular}{@{}lcccc@{}}
\toprule
Content Type & Original (KB) & Compressed (KB) & Ratio (\%) & Reduction (\%) \\ \midrule
Silence & 862 & 55 & 6.4 & 93.6 \\
Pink Noise & 938 & 538 & 57.4 & 42.6 \\
Pure Tone & 862 & 499 & 57.9 & 42.1 \\
Real Audio & 938 & 712 & 75.9 & 24.1 \\ \bottomrule
\end{tabular}
\end{table}

\begin{figure}[H]
\centering
\includegraphics[width=0.9\linewidth]{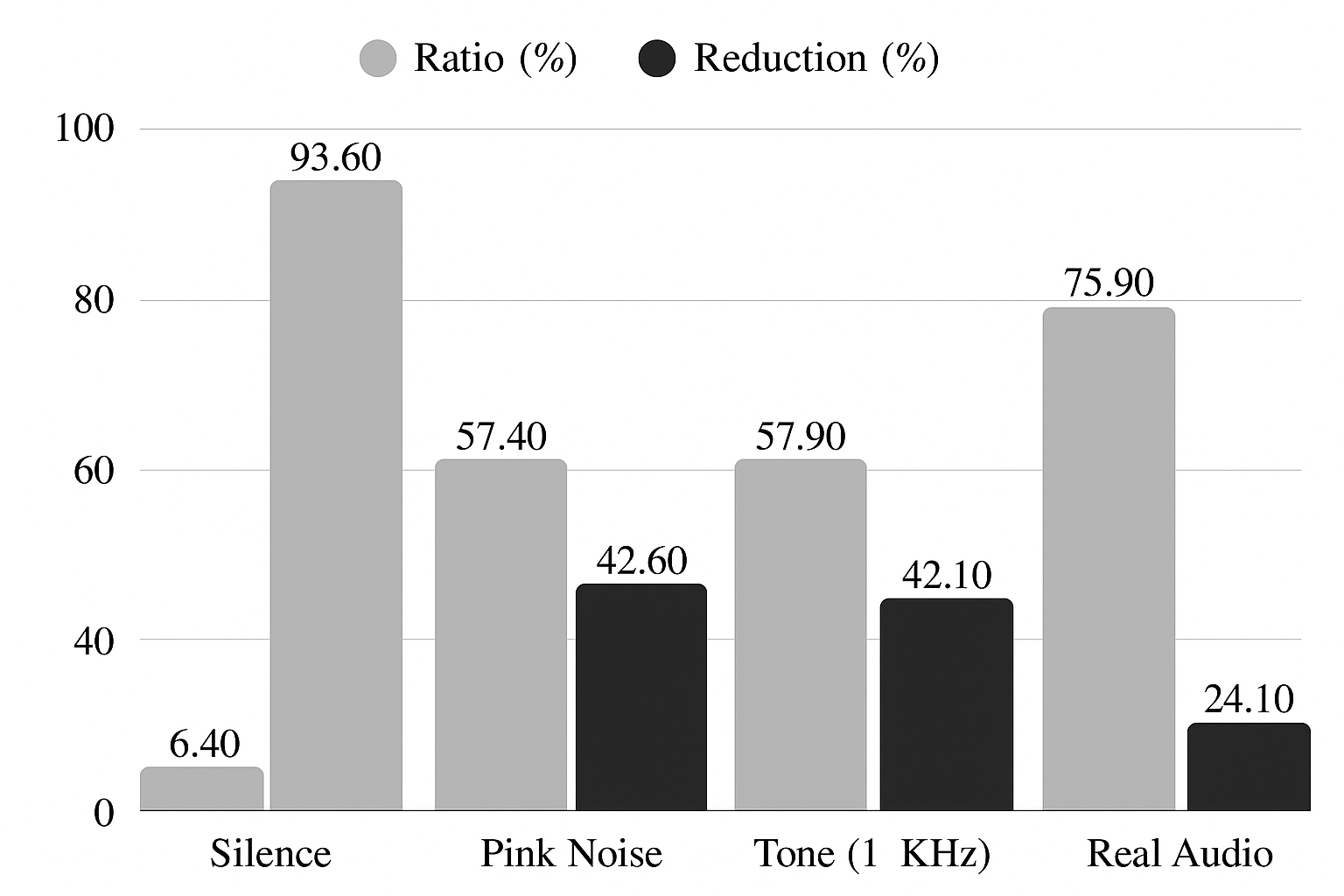}
\caption{Compression ratio and reduction percentage achieved by OBHS for various audio types.}
\label{fig:obhs_chart}
\end{figure}

\subsection{Comparison with Existing Methods}
\begin{table}[H]
\centering
\caption{Comparison with Existing Lossless Audio Codecs}
\label{tab:comparison}
\begin{tabular}{@{}lccc@{}}
\toprule
Codec & Avg. Compression (\%) & Latency (ms) & Complexity \\ \midrule
OBHS & 57.5 & 93 & Low \\
FLAC & 45.2 & 185 & Medium \\
ALAC & 46.8 & 156 & Medium \\
G.711.0 & 65.3 & 40 & Low \\ \bottomrule
\end{tabular}
\end{table}

\begin{figure}[H]
\centering
\includegraphics[width=0.9\linewidth]{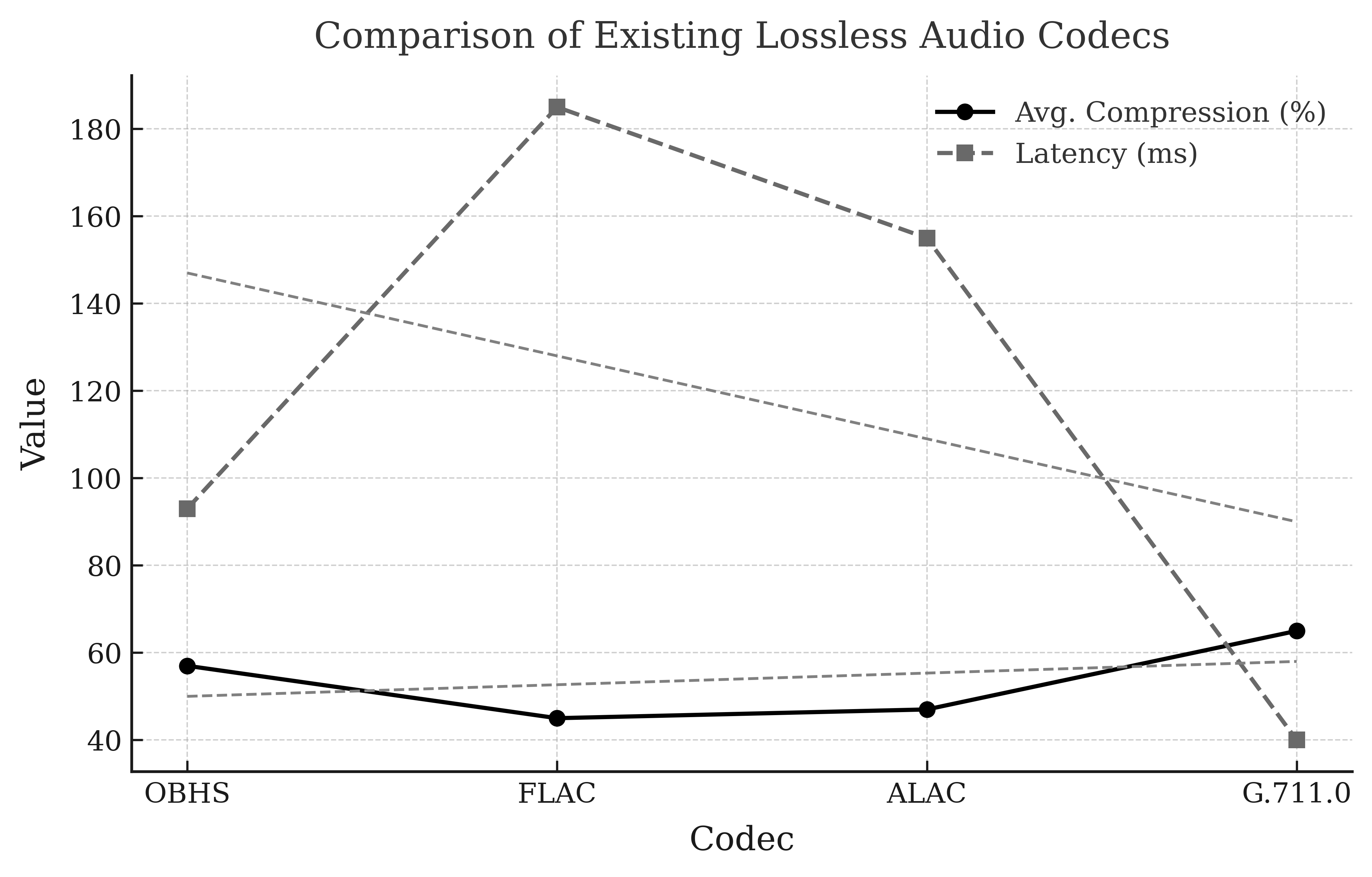}
\caption{Comparison among lossless audio codecs highlighting compression efficiency and latency.}
\label{fig:comparison}
\end{figure}

\subsection{Latency Analysis}
For a 4096-sample block at 44.1 kHz, latency $\approx$ 93 ms
 + 5 ms processing, well-suited for real-time streaming.

\section{Discussion}
The experimental results validate OBHS’s balance between efficiency and simplicity. It achieves up to 93.6\% reduction for silence and competitive results for real audio while maintaining low complexity.

\subsection{Limitations and Future Work}
\begin{enumerate}[label=(\roman*)]
\item Fixed block size may not suit all audio types.
\item No exploitation of inter-sample correlation.
\item Limited performance on highly complex signals.
\end{enumerate}
Future work will explore adaptive block sizing, hybrid entropy-predictive models, and hardware acceleration.

\section{Conclusion}
OBHS combines block-wise Huffman coding, canonical representation, and fallback mechanisms for efficient real-time audio compression. With linear complexity and deterministic latency, it provides excellent compression (up to 93.6\%) while being lightweight, making it suitable for real-time audio streaming and low-resource devices.

\section*{Acknowledgment}
The authors thank the anonymous reviewers in advance for their valuable feedback. Generative AI tools (ChatGPT, Gemini, Claude) were used only for formatting and grammatical support, not for technical contributions. The authors declare no conflict of interest.

\bibliographystyle{IEEEtran}

\end{document}